\documentclass[aps,prl,showpacs,twocolumn,superscriptaddress]{revtex4}
\usepackage{bm,color,amsmath,amssymb,mathrsfs,latexsym,graphicx,psfrag}

\newcommand{\bs}[1]{\boldsymbol{#1}}

\begin{document}
\title{Why some Iron-based superconductors are nodal while others are nodeless}
\author{Ronny Thomale}  \affiliation{Department of Physics, Princeton University, Princeton,
  NJ 08544}
\author{Christian Platt} \affiliation{Institute for Theoretical
  Physics and Astrophysics, University of W\"urzburg, Am Hubland, D
  97074 W\"urzburg}\author{Werner Hanke} \affiliation{Institute for Theoretical
  Physics and Astrophysics, University of W\"urzburg, Am Hubland, D
  97074 W\"urzburg}
\author{B. Andrei Bernevig} \affiliation{Department of Physics, Princeton University, Princeton,
  NJ 08544}


\begin{abstract}
  The symmetry of the order parameter in iron-based superconductors,
  especially the presence or absence of nodes, is still a question of
  debate.  While contradictory experiments can be explained by
  appropriately tuned theories of nodeless superconductivity in the
  iron-arsenide compounds,
  for LaOFeP all experiments clearly point to a
  nodal order parameter. We put forward a scenario that naturally
  explains the difference between the order-parameter character in
  these two sets of compounds, and use functional renormalization
  group (fRG) techniques to analyze it in detail. Our results show that, due
  to the orbital content of the electron and hole bands, nodal
  superconductivity on the electron pockets (hole pocket gaps are
  always nodeless) can naturally appear when the third hole pocket
  which lies at wavevector $(\pi, \pi)$ in the unfolded Brillouin zone
  is absent, as is the case in LaOFeP. When present, the third hole
  pocket has overwhelming $d_{xy}$ orbital character, and the
  intra-orbital interaction with the $d_{xy}$ dominated part of the
  electron Fermi surface is enough to drive the superconductivity
  nodeless (of $s^\pm$ form). However, in its absence, pair hopping,
  inter-orbital, and electron-electron intra-orbital interactions
  render the gap on the electron pockets softly nodal.
\end{abstract}
\date{\today}


\maketitle

After two years of intense research in the physics of the new
iron-based superconductors~\cite{kamihara-08jacs3296}, the symmetry of
the order parameter is still far from settled. Theoretically, 
the current opinion converged on a $s \pm$ nodeless order parameter
that changes sign between the electron (e) and hole (h) pockets, which comes
out of both the strong and the weak-coupling pictures of the
iron-based
superconductors~\cite{seo-08prl206404,mazin-08prl057003,graser-09njp025016,kuroki-08prl087004,stanev-08prb184509}.
However, most of these theories are either phenomenological
\cite{korshunov-08prb140509,chubukov-08prb134512}, or use models that
take into account only a part of the Fe orbitals present in the material. 
The most quantitative approaches give anisotropic gaps around the
electron Fermi surface which, at their smallest value, are close
but do not cross zero~\cite{wang-09prl047005}.
\begin{figure}[t]
  \begin{minipage}[l]{0.99\linewidth}
    \includegraphics[width=\linewidth]{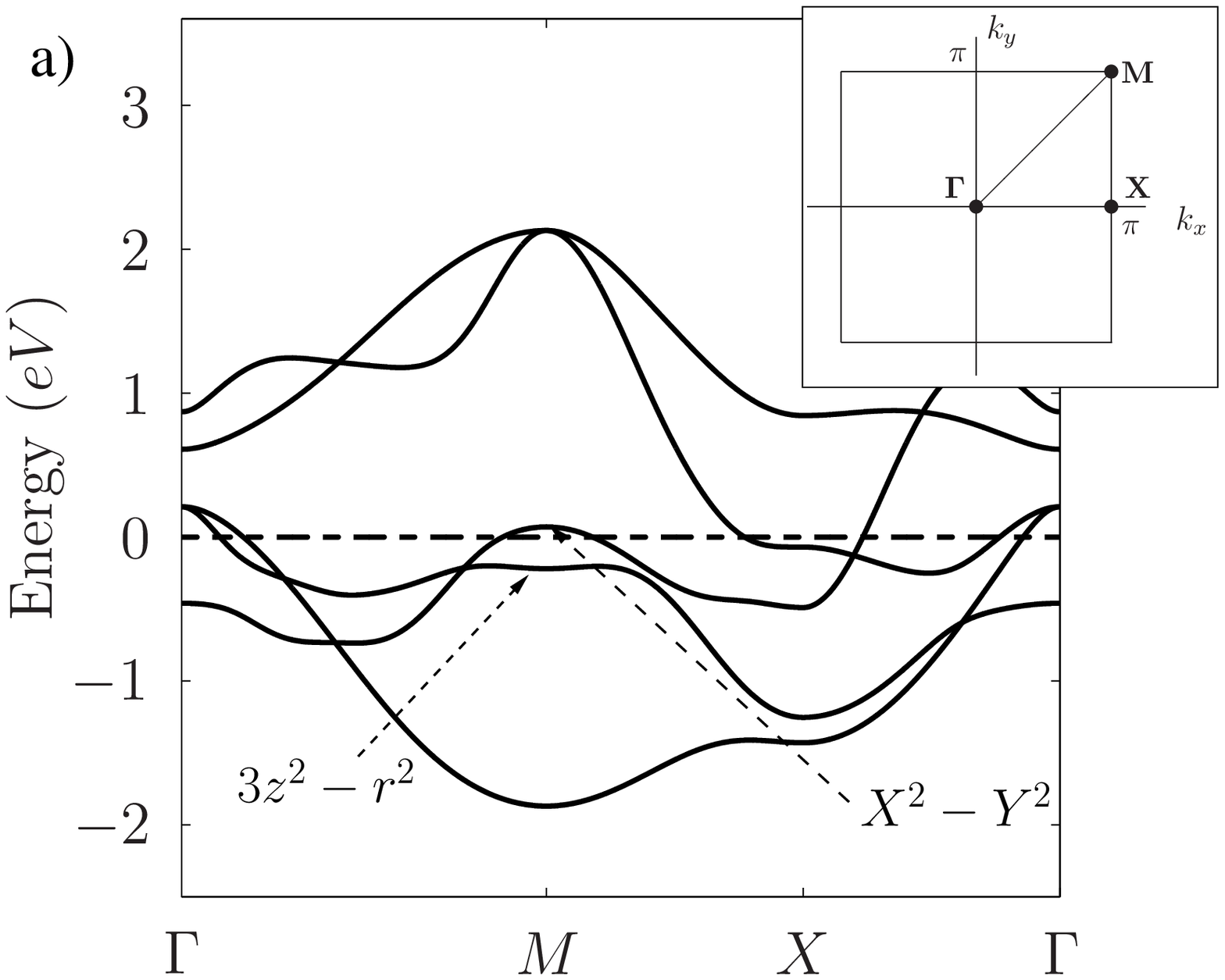}
  \end{minipage}
  \caption{2D Band structure for LaOFeAs (a) and LaOFeP (b). (Inset: Brillouin Zone). 
    The tight-binding model for LaOFeAs is contains in-plain hoppings
    up to fifth nearest-neighbors~\cite{kuroki-08prl087004}; the
    parameters are varied for LaOFeP according to the different
    pnictogen height parameters given in~\cite{kuroki-09prb224511}.
    The dashed horizontal lines denote the Fermi Level for the undoped
    compound. The electronic structure looks very similar in both
    systems. The major difference is the $d_{X^2-Y^2}$ dominated band
    crossing the Fermi level in (a), but not in (b). Being still away
    from the Fermi level, the $3z^2-r^2$ dominated band is shifted up in
    (b) compared to (a).}
\label{pic1}
\vspace{-0pt}
\end{figure}
 
The experimental situation is more controversial. In the $122$
compounds, ARPES points to the existence of nodeless isotropic gaps on
the hole Fermi surfaces \cite{wray-08prb184508,ding-08epl47001};
on the electron Fermi surface, on which ARPES also shows an isotropic gap, the data is less trustful. NMR spin relaxation time shows no
coherence peak \cite{matano-08epl57001} and exhibits a $T^3$ power-law
right below the transition temperature up to $0.1-0.2 T_c$,
reminiscent of the cuprates nodal behavior. This can be explained
by fine-tuning an $s_{\pm}$ order parameter \cite{parish-08prb144514}.
Penetration depth data on the Fe-As compounds display
power-law~\cite{hashimoto-09prl207001} and
exponential~\cite{malone-09prb140501} behavior. Thermal conductivity
shows no residual $\kappa/T$ intercept, although its in-field
dependence points to a large gap
anisotropy~\cite{tanatar-09cm0907,checkelsky-09cm0811}.  By contrast,
for LaOFeP, the penetration depth is extremely linear with
temperature, and the thermal conductivity shows a large residual
intercept, both strong characteristics of nodal
superconductivity~\cite{yamashita-09prb220509,hicks-09prl127003}. This
is even more puzzling since both materials have similar Fermi surface
topology~\cite{lu-09pc452}.

In this article, we offer an explanation for the puzzling
difference between the order parameter character in the As and P-based
compounds. Using functional renormalization group (fRG) on a $5$-band
orbital model of the iron-based superconductors with orbital
interactions, we find that the gap on the e-pockets can undergo
a nodal transition if the third hole pocket at $(\pi,\pi)$ is absent.
Both ARPES and numerical simulation data show this to be true in the case of
LaOFeP~\cite{lu-08n81,carrington-09pc459}. By using an extended
orbital model due to
\begin{figure*}[t!]
  \begin{minipage}[l]{0.85\linewidth}
    \includegraphics[width=\linewidth]{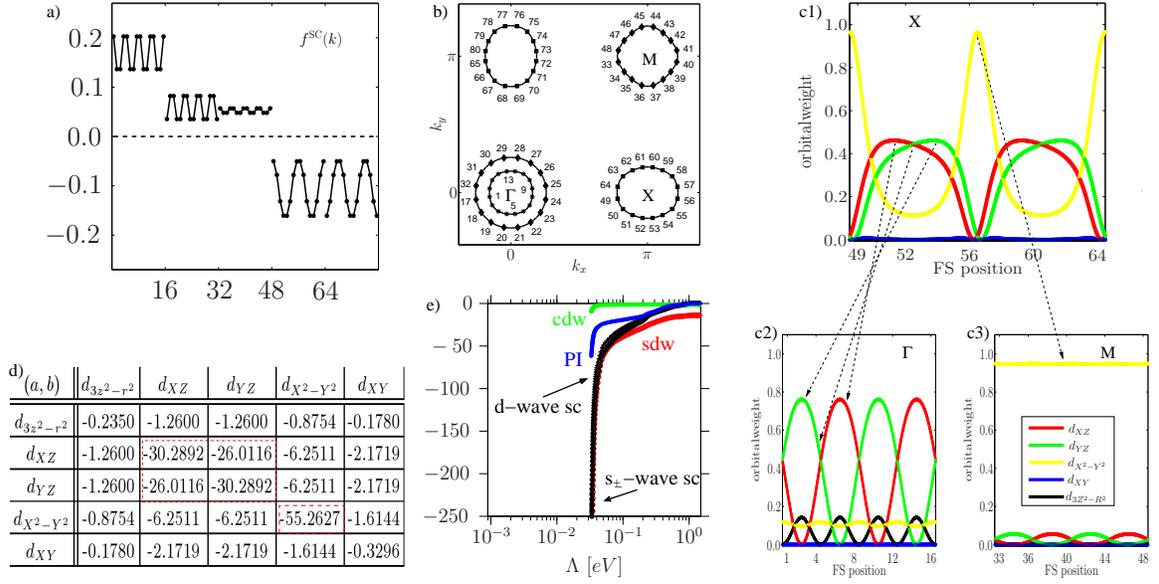}
  \end{minipage}
  \caption{5 pocket scenario for As-based compounds. a) Plot of the
    superconducting form factor gap versus the patching indices
    (momenta) shown in b).  The gap on the outer h-pocket at
    $\Gamma$ is smaller than of the inner h-pocket and of the same
    order as the $M$ pocket gap.  The gap on the e-pockets is
    very anisotropic but nodeless and of opposite sign from h-pocket gap.  c1)-c3) Orbital weight distribution on the different
    pockets (not shown is the outer h-pocket at $\Gamma$, which is
    similar to c2) shifted by 90 degrees assuring orthogonality of the
    band vectors). Dashed lines indicate most relevant scattering
    contributions for the dominating $U_1$ intra-orbital interaction.
    d) Leading pairing instability eigenvalues $c_{i,(a,b)}^{\text{SC}}(\Lambda_c)$ from the orbital decomposition
    matrix $\langle c_a^{\dagger} c_a^{\dagger} \rangle \langle c_b
    c_b \rangle $ in the Cooper channel. Most relevant weights are on
    the diagonal and off-diagonal contributions from the $d_{XZ,YZ}$
    orbitals as well as the diagonal contribution of $d_{X^2-Y^2}$. e)
    Flow of leading instability eigenvalues (charge density wave
    (CDW), Pomeranchuk instability (PI), SDW, and SC). SDW
    fluctuations are highly relevant, leading ($s^\pm$); sub-leading
    ($d$-wave) SC instabilities as well as SDW diverge in close
    proximity to each other, where the divergence scale is of $\Lambda_c
    \approx 0.08 eV$.}
\label{pic2}
\vspace{-0pt}
\end{figure*}
Kuroki {\it et al.}~\cite{kuroki-08prl087004,kuroki-09prb224511}, we
find that the third h-pocket is overwhelmingly composed of $d_{xy}$
orbital (or $d_{X^2-Y^2}$ in a $45^\circ$ rotated basis along the
Fe-As bonds).  When present, the third h-pocket at the $(\pi,\pi)$
point in the unfolded Brillouin zone (Fig.~\ref{pic2}b) scatters through intra-orbital interactions strongly with the
$d_{xy}$ part of the e-pockets around the X point to form an $s^\pm$ phase. This enhances
the already present $s^\pm$ superconductivity between the $\Gamma$
point h-pockets and the e-pockets. The gap on the e-pockets is found to be very anisotropic.  
Upon changing to the P-based structure,
the absence of the third h-pocket allows for pair hopping,
inter-orbital interactions, and electron-electron (e-e) scattering to drive
the already anisotropic gap on the e-pockets nodal.
\begin{figure*}[t]
  \begin{minipage}[l]{0.85\linewidth}
    \includegraphics[width=\linewidth]{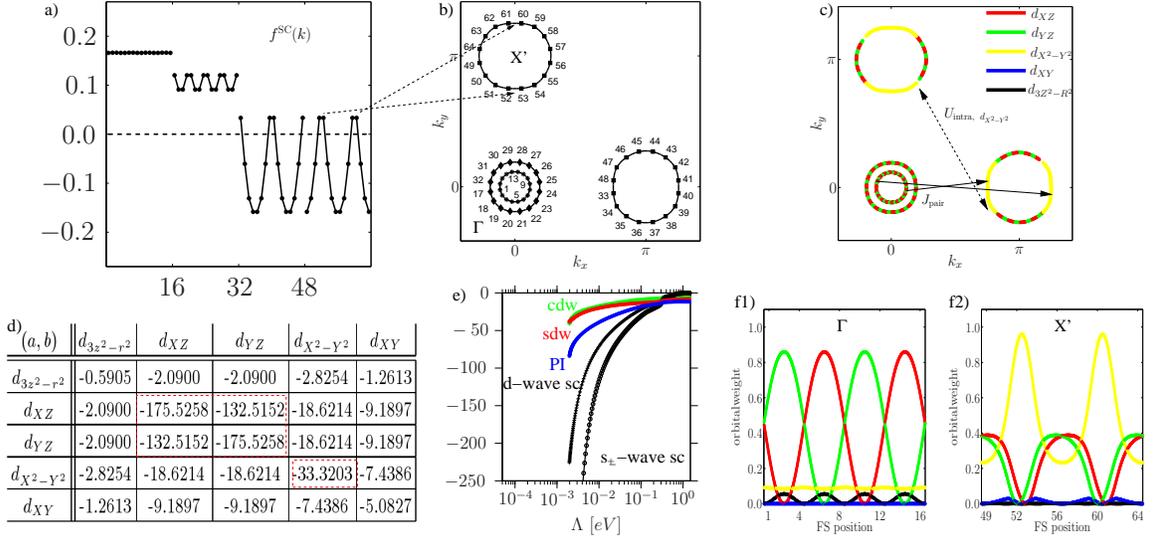}
  \end{minipage}
  \caption{4 pocket scenario for LaOFeP. a) Leading $s^\pm$
    superconducting form factor gap, with Fermi surface patches given
    in b); the $d_{X^2-Y^2}$-dominated h-pocket at $M$ is absent.
    The h-pockets at $\Gamma$ are gapped and rather isotropic, with
    a smaller gap on the outer h-pocket. The e-pockets show
    a huge anisotropy, being nodal on the inner pockets tips indicated
    by dashed arrows. c) Important processes leading to nodal
    channels. Being subleading in the presence of the $M$ hole pocket,
    $J_{\text{pair}}$ scattering processes from the $d_{XZ,YZ}$
    dominated $\Gamma$ pockets to the e-pockets and
    electron-electron intra-orbital processes between e-pockets
    now contribute to nodal propensity, with the orbital weight
    distributions shown in f1)-f2). d) The orbital decomposition
    matrix of the SC instability show increasing weight on the
    $d_{XZ,YZ}$ orbitals with a less relevant diagonal $d_{X^2-Y^2}$
    contribution. e) Flow of leading instability eigenvalues (notation
    as in Fig.~\ref{pic2}e); nodal $s^\pm$ is favored; the SDW
    fluctuations are comparably weak. The divergence scale $\Lambda_c
    \approx 0.0065 eV$ is much less than for the As-based materials.}
\label{pic3}
\vspace{-0pt}
\end{figure*}
\
We use a two-dimensional
tight-binding model developed by Kuroki {\it et al.}~\cite{kuroki-08prl087004} to describe the band structure of the
1111-type iron-based superconductors:
\begin{equation}
H_{0} = \sum_{\bs{k},s}\sum_{a,b=1}^{5}c_{\bs{k}as}^{\dagger}K_{ab}(\bs{k})c_{\bs{k}as}^{\phantom{\dagger}}.
\end{equation}
Here $c$'s denote electron annihilation operators, and $a,b$ the $5$
$d$ Fe orbitals and $s$ the spin indices.  While the main electronic
structure of P-based and As-based compounds is very similar, there are
certain important differences. Fig.~\ref{pic1} shows the band
structure of LaOFeAs and LaOFeP, where the latter is obtained by
adjusting the parameters in~\cite{kuroki-08prl087004} according to the
changed pnictogen height from As to P~\cite{kuroki-09prb224511}. In
the vicinity of the Fermi surface, the most notable difference is the
presence or absence of a broad $d_{X^2-Y^2}$ ($d_{xy}$)-dominated band
at $M=(\pi,\pi)$, in agreement with ARPES data. To account for this
difference, we use a {\it $5$ pocket scenario} for the {\it As-based}
and a {\it $4$ pocket scenario} for the {\it P-based} compounds.
We choose to compare and analyze two generic cases, with or without
the h-pocket at $(\pi,\pi)$ as corresponding to the As-based $1111$
($122$) and the P-based compounds, respectively.

The interactions in the orbital model are given by:
\begin{eqnarray}
&&H_{\text{int}}=\sum_i \left[ U_1 \sum_{a} n_{i,a\uparrow}n_{i,a\downarrow} + U_2\sum_{a<b,s,s'} n_{i,as}n_{i,bs'} \right.\nonumber \\
&& \hspace{-15pt}\left.+\sum_{a<b}(J_\text{H}\sum_{s,s'} c_{ias}^{\dagger}c_{ibs'}^{\dagger}c_{ias'}^{\phantom{\dagger}}c_{ibs}^{\phantom{\dagger}}  +J_{\text{pair}} c_{ia\uparrow}^{\dagger}c_{ia\downarrow}^{\dagger}c_{ib\downarrow}^{\phantom{\dagger}}c_{ib\uparrow}^{\phantom{\dagger}}) \right]\hspace{-4pt},
\end{eqnarray}
where $n_{i,as}$ denote density operators at site $i$ of spin $s$ in
orbital $a$. We consider intra- and inter-orbital interactions $U_1$
and $U_2$ as well as Hund's coupling $J_{\text{H}}$ and pair hopping $J_{\text{pair}}$. In what follows, we choose a physical
interaction setting dominated by intra-orbital coupling, $U_1 > U_2 >
J_\text{H} \sim J_{\text{pair}}$, and choose $U_1 = 3.5 eV, U_2 = 2.0
eV, J_{\text{H}}=J_{\text{pair}}=0.7 eV$~\cite{wang-09prl047005}.  (Even though the
interaction scales are chosen relatively high, the bare effective
interaction scale, taking into account different orbital weights, does
not exceed $\sim 2 eV$, versus a kinetic bandwidth of $\sim 5 eV$.)
This is a rather simplified picture as these scales would in general
also depend on the different orbitals.  However, for the two scenarios
representative for As- and P-based compounds, we checked that the main
features are stable under variation of these parameters. From the band
structure point of view, it should also be noted that the
$d_{3z^2-r^2}$-dominated band moves towards the Fermi level for the
P-based compound (Fig.~\ref{pic1}). However, this band only plays a
marginal role since no other relevant band has this orbital content
and hence any scattering to other bands would be governed by
sub-leading inter-orbital interactions. We defer a refined discussion
to a later stage~\cite{abinitio}.
Using functional renormalization group
(fRG)~\cite{platt-09njp055058,thomale-09prb180505,wang-09prl047005,honerkamp-01prb035109,shankar94rmp129},
we study how the renormalized interaction described by the 4-point
function (4PF) evolves under integrating high energy fermionic
modes; the flow parameter is the IR cutoff $\Lambda$ approaching the
Fermi surface, i.e. 
$V_{\Lambda}(\bs{k}_1,n_1;\bs{k}_2,n_2;\bs{k}_3,n_3;\bs{k}_4,n_4)c_{\bs{k}_4n_4s}^{\dagger}c_{\bs{k}_3n_3\bar{s}}^{\dagger}c_{\bs{k}_2n_2s}^{\phantom{\dagger}}c_{\bs{k}_1n_1\bar{s}}^{\phantom{\dagger}},$
with $\bs{k}_{1}$ to $\bs{k}_{4}$ the incoming and outgoing momenta.
Due to the spin rotational invariance of interactions, we constrain
ourselves to the $S^z=0$ subspace of incoming momenta $\bs{k}_1,
\bs{k}_2$ (and outgoing $\bs{k}_3,\bs{k}_4$) and generate the singlet
and triplet channel by symmetrization and antisymmetrization of the
4PF $V_{\Lambda}$~\cite{honerkamp-01prb035109}.  The starting
conditions are given by the kinetic bandwidth serving as an UV cutoff,
with the bare initial interactions for the 4PF. The diverging channels
of the 4PF under the flow to the Fermi surface signal the nature of
the instability, and $\Lambda_c$ serves as an upper
bound for the transition temperature $T_c$.  For a given instability
characterized by some order parameter $\hat{O}_{\bs{k}}$ (the most
important example of which is the SC instability
$\hat{O}^{\text{SC}}_{\bs{k}}=c_{\bs{k}}c_{-\bs{k}}$ in our case), the 4PF in the
particular ordering channel can be written as
$\sum_{\bs{k},\bs{p}}V_{\Lambda}(\bs{k},\bs{p})
[\hat{O}^{\dagger}_{\bs{k}}
\hat{O}^{\phantom{\dagger}}_{\bs{p}}]$~\cite{zhai-09prb064517}.
Accordingly, the 4PF in the Cooper channel can be decomposed into
different eigenmode contributions
\begin{equation}
V^{\text{SC}}_{\Lambda} (\bs{k},-\bs{k},\bs{p})= \sum_i c_i^{\text{SC}}(\Lambda) f^{\text{SC},i}(\bs{k})^* f^{\text{SC},i}(\bs{p}),
\label{decomp}
\end{equation} 
where $i$ is a symmetry decomposition index,
and the leading instability of that channel corresponds to
an eigenvalue $c_i^{\text{SC}}(\Lambda)$ first diverging under the flow of $\Lambda$.
$f^{\text{SC},i}(\bs{k})$ is the SC form factor of pairing mode $i$
which tells us about the SC pairing symmetry and hence gap structure
associated with it. In fRG, from the final Cooper channel 4PFs, this
quantity is computed along the discretized Fermi surfaces, and the
leading instability form factors are plotted in Fig.~\ref{pic2}a
and~\ref{pic3}a.

{\it As-based compounds:} For the As-based setting, we find that the
{\it $s^\pm$ instability is the leading instability} of the model at
moderate doping.  The setup resembles the situation studied
in~\cite{wang-09prl047005}, which, as an additional check, we also
studied with a more detailed tight-binding structure beyond 5th
next-nearest neighbors. We likewise find a nodeless $s^\pm$ SC leading
instability.  However, we can identify the {\it $M$ h-pocket to play a
major role} in contributing to the SDW fluctuations and to support the
full gapping of the $s$-wave as well as sign-change from hole to
electron pockets (Fig.~\ref{pic2}). In particular, we study the
orbital content in detail and analyze how the pairing instability
distributes over the different orbitals (Fig.~\ref{pic2}d). For this,
we consider the 4PF in orbital space,
\begin{eqnarray}
V^{\text{orb}}_{c,d \rightarrow a,b}&=& \sum_{n_1,\ldots,n_4 = 1}^5 \Big{\{}V_{\Lambda}(\bs{k}_1,n_1;\bs{k}_2,n_2;\bs{k}_3,n_3;\bs{k}_4,n_4)\nonumber \\
&&\times u^*_{an_1}(\bs{k}_1)u^*_{bn_2}(\bs{k}_2)u_{cn_3}(\bs{k}_3)u_{dn_4}(\bs{k}_4)\Big{\}}, 
\label{orbital}
\end{eqnarray}
where the $u$'s denote the different orbital components of the band
vectors.
The matrix shown in Fig.~\ref{pic2}d gives the leading eigenvalue
contributions of $V^{\text{SC}}_{\Lambda,
  (a,b)}(\bs{k},\bs{p})=\langle c_{\bs{k}, a}^\dagger
c_{-\bs{k},a}^\dagger \rangle \langle c_{\bs{p}, b} c_{-\bs{p}, b}
\rangle$, in the Cooper channel of~\eqref{orbital}. As
in~\eqref{decomp}, we decompose it into different form factor
contributions $\sum_i c_{i,(a,b)}^{\text{SC}}(\Lambda)
f^{i,\text{SC}}_{(a,b)}(\bs{k})^* f^{i,\text{SC}}_{(a,b)}(\bs{p})$,
where the leading eigenvalues at $\Lambda_c$ for different $(a,b)$ are
plotted in Fig.~\ref{pic2}d and~\ref{pic3}d. We observe a leading
contribution in the diagonal part of the $d_{X^2-Y^2}$ orbital, which
is strongly linked to the scattering contributions from the M h-pocket to the $d_{X^2-Y^2}$-dominated parts of the e-pockets
(Fig.~\ref{pic2}). The main scattering processes are intra-orbital
scattering between the $d_{xz}$ (or $d_{yz}$) orbitals-dominated parts
of the e- and $\Gamma$ h-pockets. These processes favor an
$s^\pm$ SC instability, as was found in~\cite{wang-09prl047005}.
However, there is another process in the case of As-based
superconductors: Along the $\Gamma \leftrightarrow X$ path, the
e-pocket has a high concentration of the $d_{X^2-Y^2}$ orbital.
This part of the e-pocket then scatters strongly with the 3rd
h-pocket at the $M$-point, which is entirely made of the
$d_{X^2-Y^2}$ orbital band. The intra-orbital repulsion on them
prefers an $s^{\pm}$-type pairing between the M h-pocket and the
e-pocket, which reinforces the already present $s^\pm$ between
the $\Gamma$ h-pockets and the $X$ ($X'$) e-pockets. With
the commonly-used assumption that $U_{\text{intra}}$ is the dominant
interaction, it then seems likely that the 3 h-pockets display a gap of
identical sign: the two $\Gamma$-pockets, which are not nested with
each other and hence can have the same gap sign and are of different
orbital content than the $M$ h-pocket. This then allows all these gaps to
have the same sign. However, the e-pockets contain contributions from all $3$
mainly relevant $d$ orbitals, and therefore scatter strongly through
$U_{\text{intra}}$ with all $3$ h-pockets, which enhances the
$s^\pm$ character of the gap. The presence of the $3$rd hole pocket is
also responsible for the strong SDW signal, being due to the fact that the nesting
wave-vector $M \leftrightarrow X$ is the same as $ \Gamma
\leftrightarrow X$. Below we will indeed see below that the absence of this
pocket weakens the SDW.

{\it P-based compounds:} Moving to the P-based compound case changes
the physical picture even qualitatively. As shown in Fig.~\ref{pic3},
we find a nodal $s^\pm$ scenario for the P-based compounds, with lower
critical divergence scale $\Lambda_c \sim T_c$ and less SDW-type
fluctuations.  The {\it absence of SDW order} in the P-based compounds is a
well-known experimental fact.  Here, the absence of the $M$ hole
pocket removes the intra-orbital scattering contribution to the
electron pockets.  This gives way to previously subleading scattering
channels like e-e scattering between the
$d_{X^2-Y^2}$-dominated parts of the e-pockets and pair hopping
from the h-pockets at $\Gamma$ to the e-pockets. They tend
to drive this part of the e-pockets nodal (Fig.~\ref{pic3}c).
As another check, we applied constant band interactions to the
As-based and P-based setups to analyze the pure Fermi surface topology
unaffected by orbital-specific interaction effects. This means that we
do not take into accout the orbital content at a Fermi surface point,
but choose a general set of couplings $g$ that is only specified by
the type of intra- or inter-pocket
scattering~\cite{platt-09njp055058,thomale-09prb180505}. We find that
the intensity of $(0,\pi)$, $(\pi,0)$ SDW fluctuations is indeed strongly
influenced by the scattering contributions $M \leftrightarrow X$.
Constant band interactions show very isotropic gaps on the e-pockets~\cite{platt-09njp055058,thomale-09prb180505}. As a consequence, the
electron gap anisotropy, rendering the electron gap nodeless or nodal,
is a phenomenon of orbital interactions, in accordance with above
elaborations. Furthermore, we varied the intra-orbital interaction
scale on the $d_{X^2-Y^2}$ orbital and the $J_{\text{pair}}$ scale in
the P-based scenario. Increasing these scales, it triggers
the propensity to form nodes on the electron pockets. Furthermore, under this
variation,  the electron pocket part of the form factor can be
considerably shifted to being more or less nodal.  

These tendencies all appear
to be consistent with experiment: we find (i) a lower divergence scale and
hence lower critical temperature, (ii) significantly enhanced low energy
density of states in the superconducting phase, and (iii) reduced SDW type
fluctuations, which, even at pronounced nesting, are insufficient to drive
the system to a leading magnetic SDW
instability~\cite{yamashita-09prb220509,hicks-09prl127003,lu-09pc452}.
The absence of the hole pocket at M also manifests itself in the
orbital decomposition of the pairing instability Fig.~\ref{pic3}d: The
diagonal contribution of $d_{X^2-Y^2}$, in comparison to the
$d_{XZ,YZ}$, is significantly reduced.

We find that the relevance of the broad band at the unfolded $M$ point
plays the major role to explain the drastic change of properties from
the As-based to the P-based 1111 compounds, rendering the former
nodeless and the latter nodal. The nodes that appear in the P-based
compounds are driven by anisotropy and are not given by any new SC
pairing symmetry, which remains of $s^\pm$ type. A more detailed work
on the technical background and broader scope of this work is in
progress~\cite{general}. 
\begin{acknowledgments}
  We thank C. Honerkamp, J. Hu, K. Kuroki, D.-H. Lee, Z. Tesanovic, A.
  Vishwanath, and F. Wang for discussions. RT is supported by a Feodor
  Lynen Fellowship of the Humboldt Foundation. RT, CP, and WH are supported by
  DFG-SPP 1458/1. BAB acknowledges support from the
  Alfred P. Sloan Foundation and Princeton University startup funds.
\end{acknowledgments}


\end{document}